\documentclass[sigconf]{acmart}
\AtBeginDocument{%
  }

\usepackage{graphicx}
\usepackage{float}
\usepackage{booktabs}
\usepackage{subcaption}
\usepackage{tabularx}
\usepackage{array}
\usepackage{hyperref}

\renewcommand\footnotetextcopyrightpermission[1]{}
\acmConference{}{}{}
\acmBooktitle{}
\acmDOI{}
\acmISBN{}

\begin{document}

\title{Enactive Drift Regulation and the Emergence Machine: A Framework for Coherent Adaptation Through Regulated Interaction}

\author{Nicholas Davis}
\email{ndavis35@gatech.edu}
\affiliation{%
  \institution{Enactive AI}
  \city{Elyria}
  \state{Ohio}
  \country{USA}
}

\renewcommand{\shortauthors}{Davis}

\begin{abstract}
Adaptive systems increasingly operate in environments characterized by persistent non-stationarity, where patterns reorganize rather than merely vary. While existing approaches such as online learning, continual learning, and adaptive filtering address performance degradation under changing data distributions, they typically treat drift as noise, error, or distribution shift to be corrected. This paper argues that such framings miss a more fundamental challenge: the loss of organizational coherence over time. We introduce Enactive Drift Regulation (EDR) as a general adaptive principle that treats drift as a regulatory signal indicating breakdowns in coherence between a system’s internal organization and its environment. Rather than treating prediction optimization or retraining as sufficient, EDR reframes adaptation as the regulation of structure—maintaining, reorganizing, or transitioning internal dynamics to sustain viable operation under change. We present the Emergence Machine as an architectural instantiation of EDR, organized around regimes, attractors, coherence measures, reorganization dynamics, and memory across regimes. By shifting the focus from error minimization to coherence regulation, this work provides a principled framework for long-duration adaptation under non-stationarity and offers a bridge between adaptive control and enactive accounts of cognition.

\end{abstract}




\maketitle
\pagestyle{plain}

\section{\textbf{Introduction: Adaptation Under Drift}}

Adaptive artificial systems are increasingly deployed in environments characterized by continuous change. From neural and physiological signals to climate dynamics, financial markets, and interactive human–machine settings, the conditions under which systems operate are rarely stationary \cite{gama2014survey,widmer1996learning}. Patterns shift, relationships reorganize, and the relevance of past experience decays unevenly over time. Yet much of contemporary machine learning and adaptive systems research remains grounded in assumptions of stationarity, or treats non-stationarity as an exceptional condition to be corrected rather than as a fundamental property of real-world environments.

Under temporally dependent streaming conditions, estimates of predictive performance can change as the data-generating process evolves, and evaluation procedures that ignore temporal structure may give misleading results \cite{zliobaite2015evaluation}. In deployed systems, such changes are commonly addressed through mechanisms such as recalibration, recent-data weighting, windowing, or retraining. These interventions are costly, brittle, and frequently insufficient. More importantly, they obscure a deeper issue: many adaptive systems are designed to learn under stable assumptions, not to regulate themselves under drift.

This paper argues that drift is not an anomaly but the norm, and that effective adaptation under such conditions requires a different computational framing. Rather than treating drift as noise, error, or distribution shift to be eliminated, we propose Enactive Drift Regulation (EDR) as a principled approach that treats drift as a first-class signal indicating breakdowns of coherence between a system and its environment. From this perspective, adaptation is not primarily a matter of updating parameters or optimizing predictions, but of regulating the organization of internal dynamics to maintain coherence over time \cite{ashby1956cybernetics}.

\subsection{\textbf{Why Stationarity Assumptions Fail}}

Many dominant approaches to learning and prediction assume that the statistical properties of the data-generating process are stable, or at least piecewise stationary \cite{ditzler2015learning}. Even online and continual learning methods often rely on implicit stationarity assumptions within bounded windows. When these assumptions hold approximately, incremental updates can track gradual change and maintain performance.

However, in many real-world domains, stationarity fails in more fundamental ways. Signals may exhibit regime changes, phase transitions, or shifts in underlying generative structure that cannot be captured by smooth parameter drift. In such cases, updating a single model—even continuously—forces incompatible patterns into a shared representational space. The result is not graceful adaptation but instability, interference, or loss of interpretability \cite{french1999catastrophic,kirkpatrick2017overcoming}.

In neural and brain–computer-interface settings, signal characteristics vary across cognitive states, tasks, individuals, sessions, and recording conditions \cite{makeig2002dynamic,millan2010combining}. In interactive systems, human behavior develops in response to the system’s actions, and the system’s subsequent inputs are therefore partly shaped by its own prior behavior \cite{suchman1987plans,licklider1960symbiosis}. Such reciprocal dependence complicates assumptions that observations are independently generated from an unchanged distribution. In all of these settings, the environment does not merely drift—it reorganizes.

Stationarity assumptions fail because they conflate two distinct phenomena: variation within a regime and transitions between regimes. While the former can often be handled through incremental learning, the latter require structural reorganization. Treating contextual transitions as undifferentiated noise can obscure the need to infer when observations are being generated under a different latent context \cite{gershman2010context}. In EDR, this insight is extended from context inference to the regulation of organizational coherence.

\subsection{\textbf{Why Retraining Is Insufficient}}

A common response to non-stationarity is retraining \cite{gama2014survey, ditzler2015learning}. When performance degrades, models are retrained on recent data, sometimes from scratch and sometimes through fine-tuning. While retraining can temporarily restore accuracy, it does not address the underlying problem and often introduces new failure modes.

First, retraining is reactive rather than regulatory. It responds to degraded performance after the fact, rather than maintaining coherence proactively as conditions change. Second, retraining often leads to catastrophic forgetting, erasing previously learned structure that may become relevant again \cite{mccloskey1989catastrophic,parisi2019continual}. Third, retraining assumes that there exists a single model configuration that can adequately capture the current environment, an assumption that fails in settings where multiple regimes coexist or recur.

More fundamentally, retraining treats adaptation as a problem of model content rather than model organization. It assumes that the primary challenge lies in fitting the right parameters, rather than in managing the structure of representations across time. As a result, retraining addresses symptoms—declining accuracy or increasing error—without providing insight into why the system’s organization has become mismatched to its environment.

In contrast, adaptive biological systems can balance stability and plasticity rather than continually rebuilding learned organization from scratch. This balance allows established patterns to persist while remaining responsive to changing conditions. This suggests that \textit{effective adaptation under drift requires regulatory mechanisms that operate at the level of system organization rather than parameter optimization alone.}

\subsection{\textbf{Why Drift Is the Norm, Not the Exception}}

Treating drift as an exceptional condition reflects a narrow view of adaptation. In open-ended, long-duration systems, drift is not a deviation from normal operation—it is the default state. Environments change, agents act within them, and those actions alter the conditions to which the system must respond \cite{clark1997being,dipaolo2005autopoiesis}. Even in seemingly static domains, long-term dynamics introduce shifts that invalidate fixed assumptions.

From this perspective, drift should be understood not merely as distribution shift, but as a breakdown of coherence between a system’s internal organization and the structure of its environment. Such breakdowns manifest as increasing instability, loss of predictive consistency, or fragmentation of internal dynamics. Crucially, these signals contain information: they indicate that the system’s current mode of operation no longer fits its context.

Recognizing drift as normative reframes the adaptive challenge. The goal is no longer to eliminate drift, but to use it as a signal for regulation. This requires mechanisms that can detect when coherence is degrading, identify the scale at which reorganization is needed, and restructure internal dynamics accordingly. Adaptation becomes an ongoing process of balancing stability and change, rather than a sequence of corrections applied to an otherwise fixed model.

\subsection{\textbf{Enactive Drift Regulation and the Emergence Machine}}

In response to these challenges, this paper introduces Enactive Drift Regulation (EDR) as a computational principle for adaptation under non-stationarity. EDR treats drift as a first-class regulatory signal and reframes adaptation as the maintenance of coherence through structural reorganization rather than error minimization alone.

We present the Emergence Machine as a concrete instantiation of this principle. Rather than operating as a monolithic predictor, the Emergence Machine maintains multiple regimes, stabilizes attractors within those regimes, and reorganizes its internal structure when coherence degrades. In doing so, it supports long-duration operation under drift without assuming stationarity or relying on continual retraining.

The remainder of this paper develops EDR as a general adaptive framework, details the architecture and dynamics of the Emergence Machine, and situates this approach in relation to existing learning and control paradigms. Together, they offer an alternative foundation for adaptive systems designed to operate where drift is not an exception, but the rule.

Within this framework, coherence is not treated as a singular metric but as a family of organizational indicators that reflect the viability of a system's current mode of operation. Example measures may include cross-scale prediction agreement, attractor persistence, regime assignment stability, structural entropy, fragmentation indices, and trajectory coherence metrics. These measures function as regulatory signals that inform when existing organization should be maintained, modified, or reorganized.

\subsection{\textbf{Contributions}}

This paper introduces Enactive Drift Regulation (EDR), a framework for understanding adaptive intelligence under conditions of prolonged non-stationarity. Rather than treating adaptation primarily as a problem of optimization, error minimization, or continual learning, EDR reframes adaptation as the ongoing regulation of organizational coherence in the presence of drift. The primary contributions of this work are:

\begin{enumerate}
\item \textbf{A regulatory perspective on adaptation.} We argue that organizational coherence is the primary variable adaptive systems should monitor and regulate under prolonged non-stationarity. This shifts the focus of adaptation from maintaining performance within a fixed organizational structure to maintaining viability through changing conditions.

\item \textbf{The concept of drift as a regulatory signal.} EDR reconceptualizes drift as an indicator of declining organizational coherence rather than merely a statistical change in data distributions or a source of prediction error. This reframing positions drift as a signal for regulation and potential reorganization.

\item \textbf{A distinction between learning and regulation.} The framework distinguishes learning, which modifies behavior within an existing organizational regime, from regulation, which determines when existing modes of organization should be maintained, reorganized, or replaced. This distinction clarifies how adaptive systems can remain viable under continual change.

\item \textbf{A multi-timescale model of adaptive organization.} EDR proposes that adaptive systems operate through nested organizational structures spanning multiple temporal scales. Coherence may therefore be monitored through indicators such as cross-scale prediction agreement, attractor persistence, regime assignment stability, structural entropy, fragmentation indices, and trajectory coherence.

\item \textbf{The Emergence Machine architecture.} We introduce the Emergence Machine as an illustrative computational realization of EDR. The architecture organizes adaptation through attractor formation, regime organization, coherence monitoring, structural reorganization, and memory across regimes, providing a concrete example of how drift-sensitive regulation can be implemented in adaptive systems.

\item \textbf{A comparative analysis of existing adaptive paradigms.} We compare EDR with continual learning, adaptive filtering, online learning, and predictive processing, illustrating that EDR addresses a distinct explanatory problem: how systems remain coherently organized as the structure of their environment changes through time.

\end{enumerate}

\subsection{\textbf{Paper Structure}}

The remainder of this paper develops Enactive Drift Regulation (EDR) and the Emergence Machine in a progressive manner. Section 2 reframes drift as a breakdown of organizational coherence rather than as noise, prediction error, or concept shift, establishing the conceptual foundations for a regulatory account of adaptation. Section 3 introduces Enactive Drift Regulation as a general adaptive principle and formalizes its core commitments, including the distinction between learning and regulation, the role of coherence as a primary adaptive variable, and the interpretation of drift as a regulatory signal.

Section 4 presents the Emergence Machine as an architectural instantiation of EDR. The architecture is described in terms of regimes, attractors, coherence measures, reorganization dynamics, memory across regimes, and multi-timescale adaptation. Particular attention is given to how Local, Regional, and Global attractors support coherence monitoring and structural reorganization under prolonged non-stationarity.

Section 5 situates EDR within the broader landscape of adaptive systems research through comparisons with continual learning, adaptive filtering, online learning, and predictive processing. These comparisons clarify how EDR addresses a distinct explanatory problem: maintaining organizational coherence as environmental structure changes through time.

Section 6 discusses the broader implications of EDR, including long-duration adaptation, robustness without rigidity, organizational interpretability, and connections to enactive approaches to cognition and adaptive intelligence. Section 7 outlines future work on coherence measurement, empirical validation, and applications across adaptive domains. Finally, Section 8 concludes by synthesizing the theoretical contribution of EDR.

\section{\textbf{Drift Reframed: From Error to Breakdown of Coherence}}

The term drift is widely used across machine learning, signal processing, and adaptive systems, yet it is rarely given a precise or unified interpretation. Depending on context, drift may refer to noise, error accumulation, distribution shift, or changes in task structure \cite{gama2014survey,quinonero2009dataset}. This section argues that these interpretations obscure a more fundamental phenomenon. Rather than treating drift as a statistical anomaly or performance degradation, we reframe it as a breakdown of coherence between a system’s internal organization and the structure of its environment. This reframing is central to Enactive Drift Regulation (EDR), as it shifts the adaptive problem from correction to regulation.

\subsection{\textbf{Drift Versus Noise}}

Drift is often conflated with noise. In signal processing and statistical modeling, noise refers to stochastic variability superimposed on an underlying signal. Noise is typically assumed to be independent, zero-mean, and uninformative with respect to the system’s structure \cite{kay1993fundamentals}. From this perspective, robustness to noise is achieved through filtering, smoothing, or averaging, all of which aim to recover a stable signal by suppressing irrelevant variation.

Drift differs fundamentally from noise. Whereas transient noise does not necessarily alter the underlying generating relation, drift reflects persistent or systematic change in that relation. Importantly, these changes are not random perturbations around a stable mean, but alterations to the mean itself—or to the structure that defines what the mean represents.

Treating drift as noise leads to inappropriate responses. Filtering strategies that suppress variability may delay detection of drift, while smoothing can actively mask the onset of structural change \cite{basseville1993detection}. From an enactive perspective, drift is not something to be eliminated, but something to be interpreted. It signals that the system’s current organization is losing its fit with the environment. Suppressing this signal prevents timely reorganization and accelerates longer-term breakdown.

\subsection{\textbf{Drift Versus Error}}

Drift is also frequently understood through the lens of error. In predictive systems, rising error is often taken as evidence that a model has become miscalibrated or outdated. The standard response is to reduce error through retraining, parameter adjustment, or model replacement.

While error and drift are related, they are not equivalent. Error is a local, outcome-based measure: it quantifies deviation between predicted and observed values at specific points in time. Drift, by contrast, is a temporal, relational phenomenon. It reflects the accumulation of mismatches across time that indicate a deeper loss of coherence.

A system may experience increasing error without drift—for example, due to transient perturbations that do not alter underlying structure. Conversely, a system may exhibit minimal short-term error while drifting structurally, particularly if it compensates locally while its internal organization becomes increasingly fragmented. In such cases, error-based triggers delay adaptation until breakdown becomes severe.

EDR therefore treats error as an insufficient signal for adaptation. While error may indicate that something is wrong, it does not reveal what kind of reorganization is needed or at what scale. Drift, understood as a breakdown of coherence, provides richer information: it indicates that the system’s current regime no longer organizes experience effectively.

\subsection{\textbf{Drift Versus Concept Shift}}

In machine learning, drift is often formalized as concept drift: changes in the mapping between inputs and outputs over time \cite{gama2014survey}. Concept drift frameworks distinguish between gradual, abrupt, and recurring shifts, and propose strategies such as windowing, ensemble models, or change-point detection to manage them \cite{bifet2007learning}.

While concept drift captures an important class of non-stationary phenomena, it remains limited in two respects. First, it presupposes a fixed task structure—typically input–output mappings—within which drift occurs. Second, it frames adaptation primarily as model selection or replacement, rather than as regulation of internal dynamics.

EDR generalizes beyond concept drift by decoupling drift from predefined tasks. In many real-world systems, especially those involving interaction, embodiment, or complex environments, there is no stable concept to drift from. Instead, the system’s engagement with the environment continuously reshapes what counts as relevant structure. Drift in such settings reflects not just changes in mapping, but \textit{changes in the organization through which experience is interpreted}.

Moreover, concept drift approaches often assume that once a shift is detected, the appropriate response is to switch to a new model. EDR, in contrast, emphasizes continuity: preserving useful structure while reorganizing where necessary. Drift does not demand wholesale replacement, but selective reconfiguration guided by coherence signals.

\subsection{\textbf{Drift as Loss of Regime Coherence}}

The core claim of EDR is that drift should be understood as a loss of regime coherence. A regime, in the present framework, is a temporally extended organization of internal dynamics that stabilizes perception, prediction, or action under particular conditions. This conception is related to work on latent contexts and context-dependent learning \cite{gershman2010context}, but extends beyond task or context inference to include attractor organization and coherence across timescales. Regimes are not static models; they are patterns of coordination that remain viable as long as the environment supports them.

Drift occurs when the coherence that sustains a regime degrades. This degradation may manifest as increasing instability, reduced predictive consistency, conflict between internal processes, or fragmentation across temporal scales. Importantly, these signals are not arbitrary—they reflect the system’s ongoing engagement with an environment that no longer conforms to the assumptions embedded in the current regime.

This reframing has significant implications. It suggests that adaptation under drift requires mechanisms for monitoring coherence, maintaining multiple regimes, and supporting transitions without collapse. It also implies that successful long-term adaptation depends less on predictive optimality and more on the system’s ability to remain organized in the face of structural change.

Drift is not noise to be filtered, error to be minimized, or concept shift to be patched. It is a signal that a system’s current organization is losing coherence with its environment. By reframing drift in this way, Enactive Drift Regulation provides a foundation for adaptive systems that can reorganize themselves under non-stationarity rather than merely reacting to its symptoms. The next section formalizes EDR as a regulatory principle and articulates how coherence, rather than accuracy alone, becomes the central variable of adaptation.

\section{\textbf{Enactive Drift Regulation (EDR)}}

This section introduces Enactive Drift Regulation (EDR) as a general computational principle for adaptive systems operating under non-stationary conditions. EDR reframes adaptation not as the optimization of predictions or the correction of error, but as the regulation of internal organization in response to drift. In doing so, it provides a principled alternative to learning-centric accounts of adaptation and establishes the theoretical foundation for the Emergence Machine described in subsequent sections.

\subsection{\textbf{Definition of Enactive Drift Regulation}}

We define Enactive Drift Regulation (EDR) as follows:

\begin{quote}
\textbf{Enactive Drift Regulation is the process by which an adaptive system maintains coherence with its environment by monitoring drift as a regulatory signal and reorganizing its active interaction mode---i.e. reconfiguring local attractor dynamics, regional regime organization, global temporal context, and the relative influence of these scales on ongoing perception, prediction, and action.}
\end{quote}

Several aspects of this definition are essential.
First, EDR treats regulation, rather than learning, as the primary adaptive act. While learning updates parameters or representations within a given organizational structure, regulation operates at the level of that structure itself. EDR is therefore concerned with when existing organization should be preserved, when it should be modified, and when it should be replaced or reconfigured.

Second, EDR is drift-driven. Drift is not treated as a residual error or performance artifact, but as an endogenous signal indicating that the system’s current mode of organization is losing its coherence. In EDR, adaptation is triggered not by failure at the output level alone, but by degradation in the internal consistency that sustains viable operation.

Third, EDR is coherence-oriented rather than objective-oriented. The goal of regulation is not to optimize a predefined cost function, but to preserve the system’s capacity to remain meaningfully coupled to its environment over time. This orientation distinguishes EDR from control-theoretic and reinforcement learning approaches that assume stable objectives or reward structures.

\subsection{\textbf{Chaos and Non-Stationarity}}

Chaotic dynamics and non-stationarity present related but distinct adaptive challenges. Chaotic systems may be generated by stable deterministic processes while exhibiting extreme sensitivity to initial conditions, causing pointwise prediction errors to grow rapidly across time \cite{lorenz1996essence}. Non-stationary systems, by contrast, undergo changes in their statistical properties, generative organization, or regime structure. A system may therefore be chaotic without being non-stationary, non-stationary without being chaotic, or subject to both conditions simultaneously.

The Emergence Machine is designed to address this combined challenge by distinguishing predictive error from structural coherence. In chaotic domains, declining pointwise accuracy does not necessarily indicate that the system has lost access to stable underlying dynamical structure. Local, Regional, and Global attractors may continue to preserve recurring structure, dynamical geometry, and regime organization beyond the horizon of exact prediction. Scale-specific structural F1 scores, attractor persistence, and cross-scale coherence can therefore provide complementary indicators of adaptive viability.

When the underlying organization itself changes, however, persistent disagreement across temporal scales, degradation of attractor stability, or rising stress may indicate genuine structural drift. Enactive Drift Regulation provides complementary organizational signals intended to help differentiate declining pointwise predictability from broader losses of regime coherence. This distinction supports adaptation through selective regulation and reorganization rather than treating every prediction error as evidence that the model must be replaced.

\subsection{\textbf{Drift as a Regulatory Signal}}

A central departure of EDR from conventional adaptive paradigms lies in how it interprets drift. Rather than viewing drift as a problem to be corrected, EDR treats it as information about the adequacy of the system’s current organization.

In predictive and learning-based systems, rising error is often the primary trigger for adaptation. However, error is a local and outcome-dependent signal: it indicates mismatch at specific points in time but provides little guidance about how the system should reorganize. Drift, as understood in EDR, is a higher-order signal. It reflects the loss of coherence across time, manifested as instability, fragmentation, or inconsistency in the system’s internal dynamics.

Crucially, drift signals are not binary. They vary in magnitude, scale, and persistence, allowing regulation to occur incrementally rather than catastrophically. Small coherence losses may call for minor adjustments or increased plasticity, while sustained or large-scale drift may require regime transition or structural reorganization. By operating on drift rather than error alone, EDR enables adaptation that is anticipatory and graded rather than reactive and disruptive.

This interpretation aligns with how adaptive biological systems respond to change. Rather than waiting for failure, such systems continuously monitor their own stability and reorganize proactively when coherence begins to degrade. EDR abstracts this principle into a computational framework suitable for artificial systems.

\subsection{\textbf{Regulation as Structural Reorganization}}

In EDR, regulation is not implemented through continuous parameter tuning alone. Instead, it involves structural reorganization of internal dynamics. Structure here refers to the patterns of coordination—regimes, attractors, coupling relations—that govern how the system interprets and responds to its environment. Structural reorganization may include:

\begin{itemize}
    \item Transitioning between regimes that encode different modes of operation

    \item Stabilizing new attractors when existing ones lose coherence

    \item Reallocating memory or influence across temporal scales

    \item Rebalancing stability and plasticity to accommodate change 

\end{itemize}
Importantly, EDR does not assume that reorganization entails discarding prior structure. On the contrary, effective regulation preserves useful organization whenever possible, enabling reactivation or reuse when conditions recur. This contrasts with retraining-based approaches, which often erase prior structure and thereby lose long-term continuity \cite{mccloskey1989catastrophic,parisi2019continual}. By separating regulation from learning, EDR clarifies a distinction that is often blurred in adaptive systems. 
\begin{quote}
\textbf{Learning modifies content within a given structure; regulation modifies the structure that makes learning meaningful.}
\end{quote}
Both processes may coexist, but they operate at different levels and on different timescales. EDR foregrounds the latter as essential for long-duration adaptation under drift.

\subsection{\textbf{Relation to Enaction}}

EDR is inspired by enactive theories of cognition, but it does not attempt to reproduce enactivism in full philosophical detail. Instead, it adopts a disciplined enactive stance focused on organizational principles relevant to adaptive systems.

From an enactive perspective, cognition is not the manipulation of internal representations but the maintenance of viable coupling between an organism and its environment. Meaning arises through action, and breakdowns in coupling signal the need for reorganization \cite{varela1991embodied,dipaolo2005autopoiesis}. EDR operationalizes this insight by treating drift as a signal of lost viability and regulation as the means by which coupling is restored.

However, EDR differs from some enactive accounts in its scope and intent. It does not make claims about subjective experience, embodiment, or consciousness. Nor does it require that artificial systems enact meaning in the human sense. Instead, EDR translates enactive ideas into a computationally tractable framework concerned with coherence, organization, and adaptation over time.

In this sense, EDR occupies a middle ground. It avoids the representational commitments of predictive and optimization-centered models \cite{clark2013whatever}, while also avoiding metaphysical claims about artificial agency. Enaction functions here as a design constraint, not a philosophical endpoint: adaptive systems must regulate their organization to remain coherently engaged with their environments.

\section{\textbf{The Emergence Machine Architecture}}

The Emergence Machine is a computational architecture designed to instantiate Enactive Drift Regulation (EDR). Rather than operating as a single predictive model optimized for accuracy, it functions as a regulator of internal organization under drift. Its purpose is not merely to produce predictions, but to maintain coherence across time by organizing, monitoring, and reorganizing its internal dynamics in response to changing conditions.

A public interactive demonstration of the Emergence Machine is available at \url{https://emergencemachine.org}, where readers can upload time-series data and explore regime detection, attractor formation, adaptive plasticity, drift regulation, and forecasting behavior in real time.

\begin{figure*}[!t]
  \centering
  \includegraphics[width=.8\textwidth]{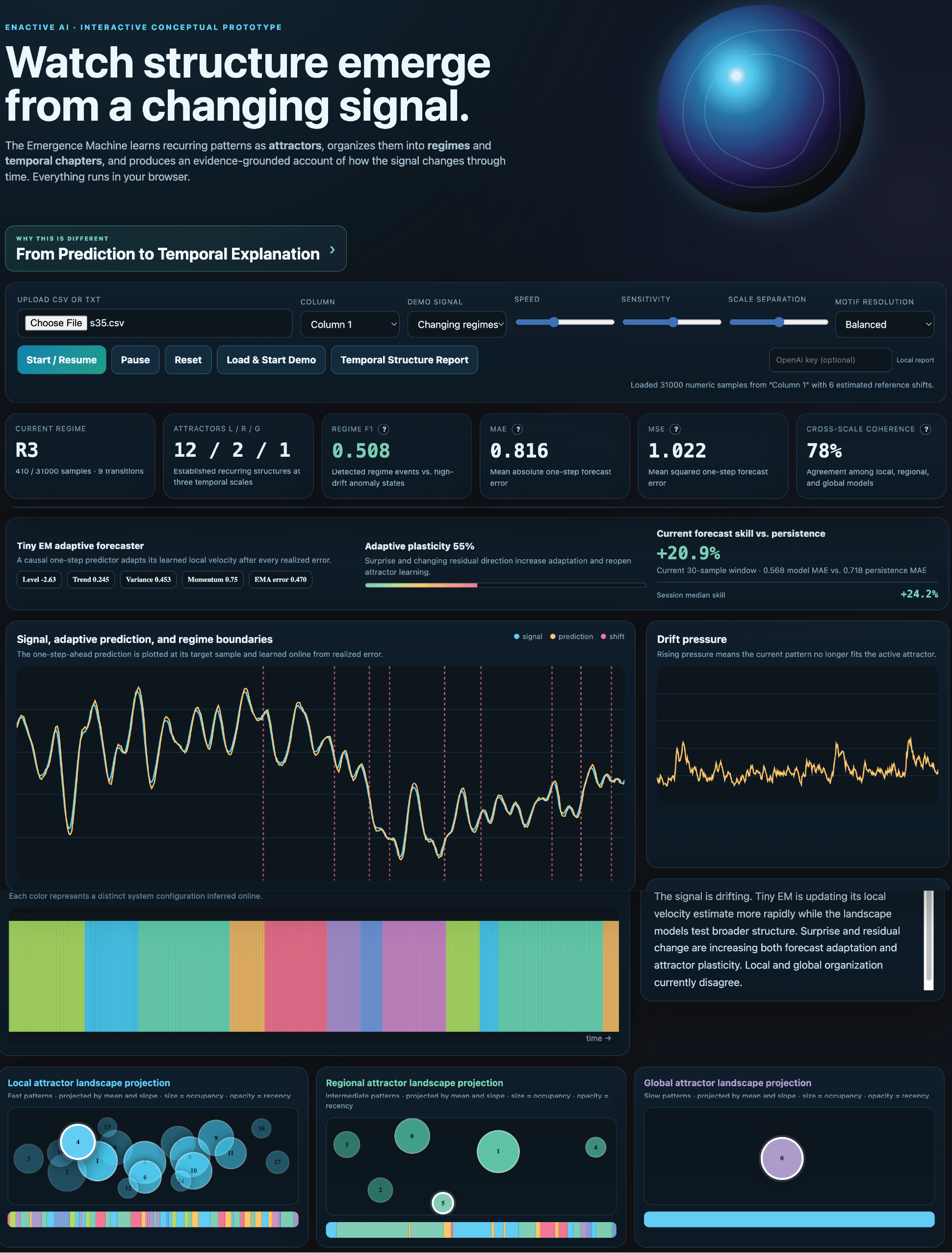}
  \caption{\textbf{The Tiny Emergence Machine browser-based interface.}
  \normalfont{The system performs fully online, one-step-ahead forecasting while organizing recurring temporal structure into Local, Regional, and Global attractors and inferred regimes. The interface exposes regime assignments, prediction error, cross-scale coherence, adaptive plasticity, persistence-relative forecast skill, drift pressure, regime boundaries, and multiscale attractor landscapes in real time. All learning, prediction, visualization, and regulation occur locally within a single lightweight webpage, providing an inspectable implementation of Enactive Drift Regulation.}}
  \label{fig:EMMVP}
\end{figure*}

\begin{figure*}[h]
  \centering
  \includegraphics[width=.9\textwidth]{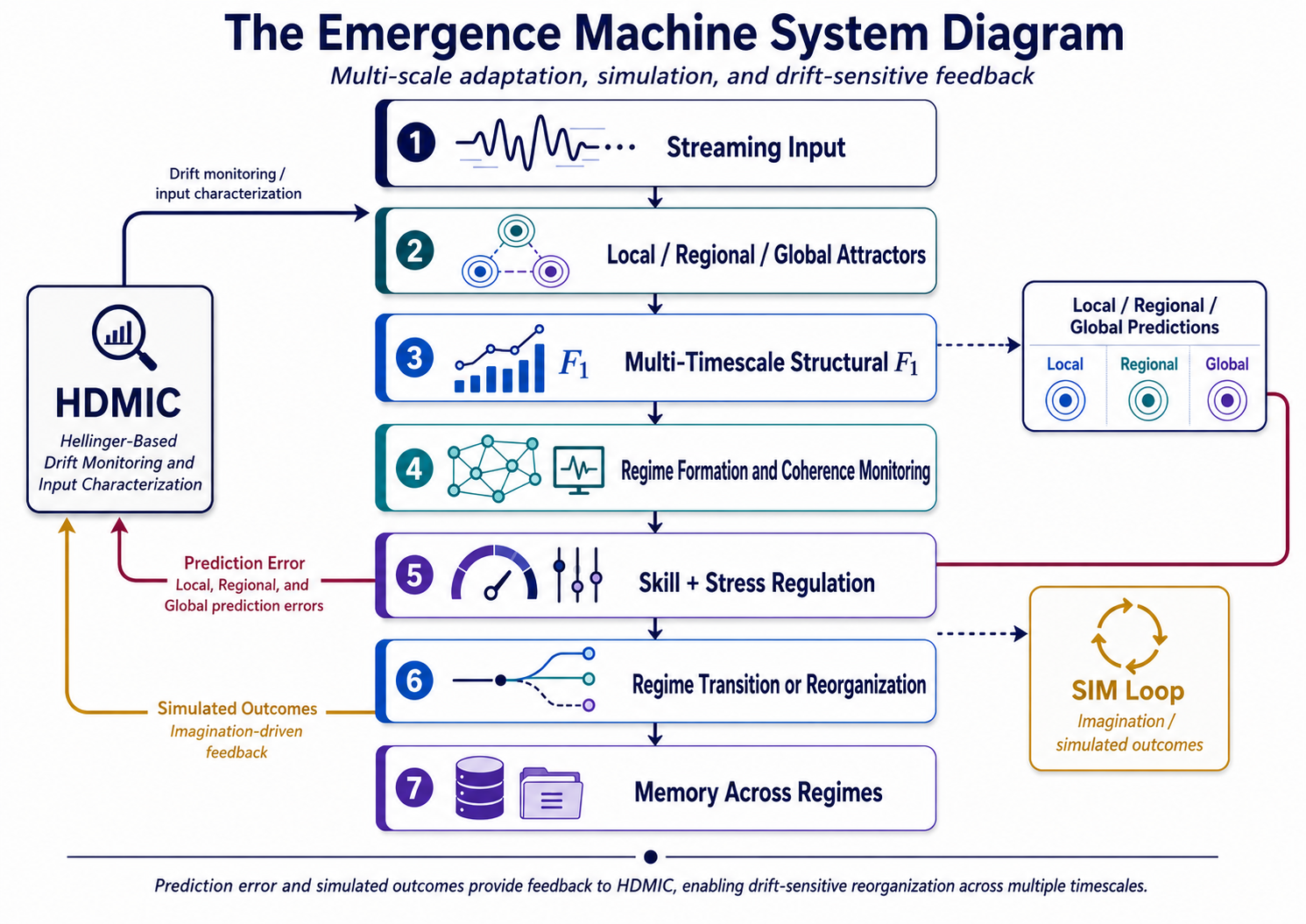}
  \caption{\textbf{The Emergence Machine system architecture.}
\normalfont{Streaming input is organized through Local, Regional, and Global attractors that capture recurring structure across nested temporal scales. Multiscale structural evaluation and coherence monitoring inform skill and stress regulation, which modulates the balance between stability and plasticity. When coherence degrades, the system may transition between regimes or reorganize its internal structure, while memory across regimes preserves previously viable configurations for later reuse. Prediction errors and simulated outcomes from the SIM loop feed back into Hellinger-Based Drift Monitoring and Input Characterization (HDMIC), forming a recurrent cycle of drift-sensitive, anticipatory regulation across timescales.}}
  \label{fig:EMDiagram}
\end{figure*}

The Emergence Machine has been developed across multiple implementation scales. Figure \ref{fig:EMMVP} presents the Tiny Emergence Machine, a compact browser-based implementation that distills selected EDR mechanisms into a lightweight, fully online forecasting system. Figure \ref{fig:EMDiagram} presents the broader architectural model, which has been implemented in an instrumented environment geared toward decision making in financial domains. This developmental architecture includes the fuller regulatory organization of attractors, regimes, structural evaluation, skill and stress dynamics, simulation, and memory across regimes. The Tiny Emergence Machine shown in Figure \ref{fig:EMMVP} illustrates how core principles of drift-sensitive, multiscale regulation can be preserved within a minimal and directly inspectable implementation.

The broader developmental architecture includes scale-specific structural evaluation measures, whereas the public Tiny Emergence Machine shown in Figure \ref{fig:EMMVP} reports a single online regime-event alignment F1 based on internally inferred high-drift and transition events.

This section describes the architectural principles of the Emergence Machine. The description is intentionally abstract: the goal is to articulate the organizational logic that supports EDR rather than to prescribe a specific implementation. The architecture is structured around five interrelated components: regimes, attractors, coherence measures, reorganization dynamics, and memory across regimes (see Figure \ref{fig:EMDiagram}). The five components describe the architecture’s organizational primitives, while Figure \ref{fig:EMDiagram} depicts the seven-stage regulatory flow through which those primitives operate.

\subsection{\textbf{Regulatory Flow Through the Emergence Machine}}

Figure~\ref{fig:EMDiagram} illustrates the regulatory flow through the Emergence Machine. Although the architecture is presented as a sequence for clarity, its components form a recurrent adaptive system in which predictions, simulated outcomes, coherence signals, and memory continually feed back into ongoing organization.

The process begins with \textbf{streaming input}. Rather than treating incoming observations as independent samples, the Emergence Machine interprets them relative to an evolving temporal history. Hellinger-Based Drift Monitoring and Input Characterization (HDMIC) evaluates changes in the incoming distribution and provides an initial characterization of whether observed variation reflects local fluctuation or a broader shift in environmental structure.

Incoming patterns are then organized through \textbf{Local, Regional, and Global attractors}. Local attractors model short-horizon recurring structure, Regional attractors capture intermediate temporal organization, and Global attractors preserve longer-horizon regularities. These nested attractor systems allow the architecture to respond to immediate changes while retaining broader structural context.

These signals contribute to \textbf{regime formation and coherence monitoring}. Regimes emerge when compatible attractors across temporal scales stabilize into a coordinated mode of operation. Coherence monitoring evaluates the persistence, alignment, and internal consistency of this organization. Drift is detected when the active regime increasingly fails to integrate incoming patterns, maintain stable attractors, or preserve agreement across timescales.

\textbf{Skill and stress} regulate how the system responds to these changing conditions. Skill represents accumulated adaptive competence at Local, Regional, and Global levels, while stress reflects regulatory pressure generated by anomaly, prediction error, uncertainty, and structural mismatch. Skill supports stable and increasingly effective engagement with recurring structure; stress indicates that the demands placed on the current organization may be exceeding its viable range. Their interaction modulates the balance between preserving existing organization and increasing plasticity.

When coherence degradation persists, the architecture supports \textbf{regime transition or structural reorganization}. Regulation may weaken an unstable attractor, shift dominance toward another regime, create a new regime, or reorganize coordination across temporal levels. The SIM loop supplements this process by generating simulated outcomes that function as an imagination component. These prospective outcomes, together with Local, Regional, and Global prediction errors, provide feedback to HDMIC, enabling the system to evaluate not only observed drift but also the possible consequences of alternative regulatory responses.

Finally, \textbf{memory across regimes} preserves previously viable organizational patterns. Rather than storing only historical observations or fixed parameter states, the architecture retains configurations of attractors, regimes, and cross-scale relations that have supported coherent operation in the past. This allows prior regimes to be reactivated when conditions recur and reduces the need to relearn useful structure from the beginning.

Taken together, this flow establishes a closed regulatory cycle. Streaming input is interpreted through nested learned structures; scale-specific structural evaluation, prediction error, skill, and stress characterize the viability of the current regime; simulated outcomes support anticipatory evaluation; and drift-sensitive feedback guides maintenance, transition, or reorganization. The Emergence Machine therefore does not merely update a predictor as new data arrive. It continually regulates the organization through which prediction, learning, and adaptation remain coherent over time.

\subsection{\textbf{Regimes: Structured Modes of Operation}}

At the highest level, the Emergence Machine organizes its operation into regimes. A regime is a temporally extended mode of internal organization that stabilizes perception, prediction, or action under a particular set of conditions. Regimes are not static models, nor are they tied to predefined tasks. Instead, they are patterns of coordination that remain viable as long as they maintain coherence with incoming signals.

Each regime encodes a distinct way of structuring experience. Different regimes may correspond to different signal characteristics, behavioral contexts, or environmental conditions. Importantly, regimes can coexist: the Emergence Machine does not assume that only one regime is valid at a time. Instead, it maintains a repertoire of regimes and regulates which ones are active, dominant, or latent.

This regime-based organization enables the system to avoid forcing incompatible patterns into a single representational space. Rather than continuously adapting one model to fit all conditions, the Emergence Machine can stabilize multiple modes of operation and transition between them when coherence degrades.

\subsection{\textbf{Attractors: Stabilizing Internal Dynamics}}

Within each regime, the Emergence Machine maintains attractors—stable patterns of interaction dynamics that organize ongoing processing. Attractors may take many forms depending on the implementation: recurrent trajectories, stable parameter configurations, or persistent relational patterns. What defines an attractor is not its mathematical form, but its functional role in stabilizing behavior over time \cite{kelso1995dynamic}.

Attractors provide local coherence. They enable the system to remain consistent in the face of variation and noise, preventing overreaction to transient perturbations. At the same time, attractors are not immutable. Their stability is continually assessed through coherence measures, and they may weaken, shift, or dissolve as conditions change \cite{haken1983synergetics}.

Crucially, attractors are regime-specific. An attractor that is coherent within one regime may be incoherent in another. This separation allows the Emergence Machine to preserve useful structure without forcing it to generalize beyond its viable domain.

\subsection{\textbf{Coherence Measures: Monitoring Viability}}

Central to EDR is the ability to assess whether current organization remains viable. The Emergence Machine does this through coherence measures—signals that quantify the internal consistency and stability of ongoing dynamics.

Coherence measures are not limited to prediction error. While error may contribute, coherence is fundamentally relational and temporal. It reflects how well internal processes agree with one another across time, how stable attractors remain under perturbation, and whether regime-level organization continues to organize incoming signals effectively. Examples of coherence indicators may include:

\begin{itemize}
    \item Stability of internal trajectories

    \item Consistency of regime assignments across time

    \item Alignment between short-term and long-term dynamics

    \item Rising internal entropy or fragmentation

\end{itemize}
The precise form of these measures is implementation-dependent. What matters is their role: coherence measures serve as regulatory signals that inform whether the current organization should be maintained, adjusted, or reorganized. In this sense, they operationalize drift as a first-class signal.

\subsection{\textbf{Reorganization Dynamics: Regulating Structure}}

When coherence degrades beyond tolerable bounds, the Emergence Machine engages reorganization dynamics. Reorganization is not a binary reset or wholesale retraining. Instead, it is a graded process that adjusts internal organization at the appropriate scale. Reorganization dynamics may include:

\begin{itemize}
    \item Weakening or dissolving unstable attractors

    \item Shifting dominance between regimes

    \item Creating new regimes when existing ones fail

    \item Reallocating influence across temporal scales

\end{itemize}
Importantly, reorganization is selective and conservative. The goal is not to maximize change, but to restore coherence with minimal disruption. Where possible, existing structure is preserved and repurposed rather than discarded. This supports continuity and reduces catastrophic forgetting \cite{mccloskey1989catastrophic}.

Reorganization dynamics embody the core principle of EDR: adaptation through regulation of structure rather than through continuous optimization alone. They enable the system to remain organized under drift without assuming stationarity or relying on external intervention.

\subsection{\textbf{Memory Across Regimes: Continuity Without Freezing}}

Long-duration adaptation requires memory, but not the kind of memory that freezes a system into outdated configurations. The Emergence Machine therefore maintains memory across regimes—a form of structural memory that preserves the availability of past organizational patterns without forcing their continued dominance.

This memory allows regimes to be reactivated when conditions recur, supporting recurrence and cyclic environments. It also enables the system to compare current conditions with prior organizational states, informing reorganization decisions.

Crucially, memory across regimes is not a simple archive of past data or parameters. It is a memory of organizational solutions: ways of structuring dynamics that were once coherent. By retaining this memory, the Emergence Machine avoids relearning from scratch and supports adaptive reuse.

\subsection{\textbf{Architectural Summary}}

Taken together, these components form an architecture that instantiates Enactive Drift Regulation:

\begin{itemize}
    \item Regimes provide multiple, structured modes of operation.

    \item Attractors stabilize behavior within regimes.

    \item Coherence measures monitor the viability of current organization.

    \item Reorganization dynamics adjust structure in response to drift.

    \item Memory across regimes preserves continuity across change.

\end{itemize}
The Emergence Machine is therefore not best understood as a predictor, classifier, or controller. It is a \textbf{regulator of predictive organization}—a system designed to remain coherently organized as its environment changes. By separating regulation from learning and treating drift as a signal rather than a defect, the Emergence Machine provides a concrete instantiation of EDR suitable for long-duration, non-stationary domains. 

The Emergence Machine may be applicable to domains in which continuous regulation is more important than static prediction. For example, in critical care, physiological monitoring systems often operate under non-stationary conditions in which patients transition among dynamically evolving states. Rather than predicting exact treatment parameters, the Emergence Machine could identify transitions between physiological regulatory regimes, providing clinicians with continuously updated estimates of patient state that may support adaptive intervention strategies. Following substantial validation, EDR-based architectures may eventually support research in domains such as EEG monitoring, ECG analysis, glucose regulation, industrial process control, and autonomous robotics.

The next section situates this architecture in relation to existing adaptive paradigms, clarifying how EDR and the Emergence Machine address a different class of problems than conventional learning-based approaches.

\section{\textbf{EDR in Relation to Existing Adaptive Frameworks}}

Enactive Drift Regulation (EDR) does not compete directly with existing adaptive learning paradigms on performance metrics such as accuracy, convergence speed, or sample efficiency. Instead, it addresses a different class of adaptive problem: how systems can remain coherently organized over time when the structure of their environment changes in ways that violate stationarity assumptions. This section situates EDR in relation to several influential approaches—continual learning, adaptive filtering, online learning, and predictive processing—clarifying both points of contact and fundamental differences.

\begin{table*}[t]
\centering
\caption{\textbf{Comparison of Enactive Drift Regulation with established adaptive paradigms.}
\normalfont{Continual learning, adaptive filtering, online learning, and predictive processing primarily seek to preserve or improve performance as conditions change. Enactive Drift Regulation (EDR) instead treats drift as a signal of declining organizational coherence and regulates regimes, attractors, and memory across timescales.}}
\label{tab:edr_comparison}

\renewcommand{\arraystretch}{1.35}
\setlength{\tabcolsep}{5pt}

\begin{tabularx}{\textwidth}{
    >{\raggedright\arraybackslash}p{2.25cm}
    >{\raggedright\arraybackslash}X
    >{\raggedright\arraybackslash}X
    >{\raggedright\arraybackslash}X
    >{\raggedright\arraybackslash}X
}
\toprule
\textbf{Paradigm} &
\textbf{Primary Adaptive Objective} &
\textbf{Interpretation of Drift} &
\textbf{Typical Adaptive Mechanism} &
\textbf{Organizational Assumption} \\
\midrule

\textbf{Continual Learning}
&
Acquire new knowledge while preserving performance on previously learned tasks.
&
A transition between tasks or data distributions that may produce catastrophic forgetting.
&
Rehearsal, regularization, parameter protection, modularization, or memory replay.
&
Knowledge is organized around identifiable tasks or sequential learning episodes. \\

\textbf{Adaptive Filtering}
&
Track a changing signal while maintaining accurate state estimates.
&
Gradual variation in the parameters of an otherwise stable generative process.
&
Continuous parameter estimation, covariance updating, and recursive correction.
&
The basic form of the model remains valid while its parameters change. \\

\textbf{Online Learning}
&
Incrementally optimize predictive performance as streaming observations arrive.
&
A changing data distribution that requires faster updating, discounting, or revised weighting.
&
Learning-rate adjustment, sliding windows, sample reweighting, and incremental optimization.
&
A single objective and model organization remain sufficiently stable to support continual updating. \\

\textbf{Predictive Processing}
&
Reduce prediction error or variational free energy through continual model refinement.
&
Prediction error indicating mismatch between a hierarchical generative model and incoming signals.
&
Belief updating, parameter adjustment, model selection, and action that reduces prediction error.
&
Adaptation generally occurs within a unified hierarchical generative organization. \\

\midrule

\textbf{Enactive Drift Regulation}
&
Maintain coherent organization and viable operation as environmental structure changes.
&
A multi-timescale regulatory signal indicating loss of coherence within or across regimes.
&
Coherence monitoring, multiscale attractor and regime organization, structural reorganization, and memory across regimes.
&
Multiple regimes and attractors may coexist, recur, and reorganize across nested temporal scales. \\

\bottomrule
\end{tabularx}
\end{table*}

\subsection{\textbf{EDR and Continual Learning}}

Continual learning frameworks aim to enable systems to acquire new knowledge without catastrophically forgetting prior knowledge. Techniques such as rehearsal, regularization, and modularization seek to preserve past task performance while learning new tasks sequentially \cite{kirkpatrick2017overcoming, parisi2019continual,mccloskey1989catastrophic}.

While continual learning addresses an important problem,  approaches often rely on a task-centric framing: the environment is decomposed into a sequence of tasks, each of which can be learned, stored, and protected. Drift is treated as a transition between tasks, and success is measured by retention of performance across that sequence.

EDR departs from this framing in two key ways. First, it does not assume that environmental change can be decomposed into discrete tasks. In many real-world domains, especially continuous time-series and interactive settings, change is gradual, overlapping, and recurrent rather than task-bounded \cite{gama2014survey}. Second, EDR is not primarily concerned with preserving task performance, but with maintaining organizational coherence.

From an EDR perspective, catastrophic forgetting is a symptom rather than the core problem. The deeper issue is that forcing incompatible regimes into a single organizational structure creates internal incoherence. By explicitly maintaining multiple regimes and regulating transitions between them, EDR avoids the need to freeze parameters or protect representations. Forgetting becomes a controlled reallocation of dominance rather than a failure to retain content.

Recent work has broadened the scope of continual learning beyond catastrophic forgetting to include resource-efficient adaptation, representation stability, online updating, and deployment under continuously evolving data streams. Contemporary surveys emphasize that the field is increasingly concerned with balancing stability and plasticity while enabling long-term adaptation in realistic, non-stationary environments rather than isolated benchmark tasks \cite{Wang2024ContinualSurvey,Zhou2024PTMContinualSurvey}.

From the perspective of Enactive Drift Regulation, these developments represent an important evolution of continual learning but also highlight a remaining conceptual gap. Even as modern continual learning increasingly addresses dynamic environments, adaptation is still commonly framed in terms of preserving learned knowledge while incrementally optimizing predictive performance \cite{Wang2024ContinualSurvey,Zhou2024PTMContinualSurvey}. EDR instead treats organizational coherence itself as the primary adaptive variable. Learning remains important, but it is situated within a higher-order regulatory process that determines when existing regimes should be maintained, reorganized, or replaced in response to persistent structural drift. This distinction positions EDR as complementary to contemporary continual learning research while addressing a different explanatory objective: sustaining coherent organization rather than solely preserving accumulated knowledge.

\subsection{\textbf{EDR and Adaptive Filtering}}

Adaptive filtering techniques, such as Kalman filters and their variants, are designed to track changing signals by continuously updating estimates in response to new data \cite{kalman1960new, haykin2002adaptive}. These methods excel in environments where changes are smooth and can be captured by incremental parameter updates.

However, many conventional adaptive-filtering approaches assume that the form of the model remains valid as parameters change \cite{arenas2015combinations}. Drift is treated as gradual variation around a stable generative process, and adaptation consists of tracking that variation as efficiently as possible.

EDR addresses a different scenario: environments in which the structure of the generative process itself changes. In such cases, no amount of parameter tuning within a fixed filter structure is sufficient. The problem is not tracking, but reorganization.

Where adaptive filtering emphasizes responsiveness and smoothness, EDR emphasizes regime viability. A regime may be locally stable and still globally incoherent. EDR therefore introduces coherence monitoring and regime transition as first-class mechanisms, enabling adaptation through structural change rather than continuous estimation alone.

\subsection{\textbf{EDR and Online Learning}}

Online learning frameworks update models incrementally as data arrives, often under assumptions of bounded regret or slowly changing distributions \cite{cesabianchi2006prediction, shalevshwartz2012online}. These approaches are well suited to streaming data and non-stationary environments where adaptation must occur continuously.

Despite their flexibility, online learning methods typically optimize a single objective function over time. Drift is handled by adjusting learning rates, discounting older data, or sliding windows to emphasize recent observations. While these techniques can improve responsiveness, many online learning approaches primarily focus on updating model parameters, weights, or data weighting strategies rather than explicitly regulating organizational coherence.

EDR differs by decoupling adaptation from optimization. Rather than continuously chasing an objective that itself may be drifting, EDR treats coherence as the primary variable to regulate. When coherence degrades, the appropriate response may not be faster learning, but structural reconfiguration—activating a different regime, stabilizing a new attractor, or reorganizing memory across timescales.

In this sense, EDR complements rather than replaces online learning. Learning may occur within regimes, but regulation determines which regime is active and when learning should be emphasized or suppressed.

\subsection{\textbf{EDR and Predictive Processing}}

Predictive processing and active inference frameworks model cognition as the minimization of prediction error or variational free energy \cite{friston2010freeenergy,clark2013whatever}. These approaches have been influential in both cognitive science and machine learning, emphasizing hierarchical generative models and continual updating.

EDR shares with predictive processing an emphasis on coherence and stability rather than static correctness. However, it diverges in how adaptation is conceptualized. Predictive processing frameworks are often interpreted as treating prediction error minimization as a central driver of adaptation, with reorganization occurring implicitly through parameter updates or model selection.

In contrast, EDR treats drift—not error—as the central regulatory signal. Error may increase for many reasons, including noise or transient perturbations, without indicating a need for structural change. Drift, as loss of regime coherence, directly signals when the system’s current organization is no longer viable. 

Predictive processing generally explains adaptation through the refinement, selection, or weighting of generative models in response to prediction error. EDR instead foregrounds organizational coherence and regime viability as explicit regulatory variables.

Across these comparisons, a consistent distinction emerges. These paradigms are not incapable of representing organizational change. Rather, EDR differs in making organizational coherence itself the primary object of regulation. Continual learning, adaptive filtering, online learning, and predictive processing all ask some version of the question:

How can a system improve or maintain performance as conditions change?

EDR asks a different question:

How can a system remain coherently organized as the structure of its environment changes?

This shift in focus—from performance to coherence, from learning to regulation, from error to drift—defines the contribution of Enactive Drift Regulation. EDR does not replace existing adaptive methods; it reframes the problem they implicitly assume. In domains where non-stationarity is the norm and structural change is unavoidable, EDR offers a principled foundation for building systems that can endure.

\section{\textbf{Discussion: What Enactive Drift Regulation Enables}}

The preceding sections have argued that drift is not an exceptional failure mode but a defining condition of real-world adaptive systems, and that Enactive Drift Regulation (EDR) offers a principled response by reframing adaptation as the regulation of coherence rather than the optimization of performance. This discussion elaborates what this reframing enables in practice and why it matters for the design of adaptive systems intended to operate over extended durations in non-stationary environments.

\subsection{\textbf{Long-Duration Adaptation Without Freezing}}

A central consequence of EDR is its support for long-duration adaptation under conditions of continual drift. Performance estimates for streaming classifiers can deteriorate or become unreliable as temporal dependence and changing distributions accumulate \cite{zliobaite2015evaluation}. Traditional responses—retraining, recalibration, parameter freezing, or constrained updates—can temporarily restore stability, but often do so at the cost of flexibility and continuity \cite{kirkpatrick2017overcoming}.

EDR offers an alternative by treating drift as a regulatory signal rather than a terminal failure. Coherence is monitored continuously, allowing regimes and attractors to be maintained, modified, or reorganized before breakdown becomes catastrophic. Stability is achieved through organizational structure, while plasticity is preserved through regulated reorganization across timescales.

This approach reframes robustness not as resistance to change, but as the capacity to change without losing organization. Systems remain stable because they can reorganize when necessary rather than attempting to suppress variation entirely. In these settings, adaptation must preserve coherence across hours, days, or months rather than optimizing for short-term performance alone.

\subsection{\textbf{Coherence as a Design Principle}}

A broader implication of EDR is that coherence may serve as a primary design principle for adaptive systems. Contemporary approaches often organize adaptation around objectives such as prediction accuracy, reward maximization, error minimization, or task performance. While these objectives remain valuable, they provide limited guidance when environmental structure itself changes.

EDR proposes that adaptive systems should monitor and regulate organizational coherence directly. From this perspective, measures such as cross-scale prediction agreement, attractor persistence, regime assignment stability, structural entropy, fragmentation indices, and trajectory coherence become important not merely as diagnostic tools but as regulatory variables. These indicators provide information about whether the system's current mode of organization remains viable under changing conditions.

This shift has practical consequences for system design. Rather than asking how a model can continue optimizing within a fixed structure, EDR asks how the structure itself should be maintained, reorganized, or replaced as drift accumulates. Adaptation therefore becomes a problem of preserving coherence across time rather than maximizing performance at isolated moments.

Viewed in this way, coherence functions as an organizational counterpart to accuracy. Accuracy evaluates the quality of particular outputs, whereas coherence evaluates the viability of the organizational processes that generate those outputs. EDR suggests that long-duration adaptive intelligence may depend less on optimizing predictions than on sustaining coherence across changing conditions.

\subsection{\textbf{Interpretability Through Regimes}}

EDR also has implications for interpretability, an increasingly important concern in adaptive and learning systems. Many high-performing models, particularly deep learning architectures, are difficult to interpret because their internal organization evolves continuously and opaquely \cite{lipton2018mythos}.

By organizing behavior into regimes and attractors, the Emergence Machine provides a natural unit of interpretability. Regimes correspond to distinct modes of operation that can be inspected, compared, and related to environmental conditions. Transitions between regimes signal meaningful changes in how the system is engaging with its environment, rather than arbitrary parameter drift.

This form of interpretability is not explanatory in the sense of feature attribution or causal graphs, but it is organizationally transparent. It allows observers to understand what kind of situation the system believes it is in and how that belief changes over time. Such transparency is particularly valuable in long-duration systems, where understanding why behavior shifts is as important as the behavior itself.

\subsection{\textbf{Alignment with Embodied and Enactive Accounts}}

EDR draws on embodied and enactive accounts in which cognition is sustained through ongoing engagement with the environment rather than performed as an isolated computation over fixed representations \cite{varela1991embodied,dipaolo2005autopoiesis}. From this perspective, breakdowns in coupling are not merely errors to be corrected but indications that the organization supporting viable engagement may need to change. EDR translates this insight into a design principle by treating drift as loss of coherence and regulation as the reorganization of the system--environment relation.

This alignment remains deliberately limited. EDR does not attribute consciousness, subjective experience, or full biological embodiment to artificial systems. Instead, enaction functions as an organizational constraint: adaptive systems should monitor and regulate the conditions that sustain coherent coupling over time. This position allows artificial systems to exhibit enactively informed organization without making claims about enactive phenomenology, providing a practical bridge between cognitive theory and adaptive system design \cite{clark2013whatever}.

\section{\textbf{Future Work}}

The present work introduces Enactive Drift Regulation (EDR) as a theoretical framework and the Emergence Machine as an illustrative architectural instantiation. Future work will focus on formalizing coherence as a measurable adaptive variable. While this paper has identified candidate indicators including cross-scale prediction agreement, attractor persistence, regime assignment stability, structural entropy, fragmentation indices, and trajectory coherence, additional work is required to develop principled coherence metrics, establish relationships among these measures, and determine how coherence can be quantified consistently across domains.

A second direction concerns empirical evaluation. Future studies should compare EDR-based systems against continual learning, online learning, adaptive filtering, and predictive processing approaches under conditions of prolonged non-stationarity. Such comparisons would help determine whether coherence-based regulation improves long-duration adaptation, robustness to regime shifts, recovery from drift, and organizational stability across changing environments. Benchmark evaluation on synthetic and real-world time-series datasets represents an important next step toward validating the framework.

Finally, EDR opens opportunities for applications beyond time-series prediction. Potential domains include human--AI interaction, co-creative systems, adaptive robotics, physiological monitoring, and long-duration autonomous agents. In particular, systems such as the Emergence Machine provide an opportunity to investigate how coherence-based regulation can support sustained collaboration between humans and artificial partners. More broadly, EDR suggests a research program centered on understanding how adaptive systems can regulate organizational coherence across time, enabling learning, reorganization, and continued viability under conditions of continual change.

\section{\textbf{Conclusion}}

This paper introduced Enactive Drift Regulation (EDR) as a general adaptive principle for systems operating under persistent non-stationarity, reframing drift not as noise, error, or an exceptional condition to be corrected, but as a regulatory signal indicating loss of coherence between a system's internal organization and its environment. Whereas continual learning, online learning, adaptive filtering, and predictive processing primarily address how systems maintain or improve performance as conditions change, EDR addresses how systems remain coherently organized when environmental structure itself reorganizes. This shift from performance to coherence, and from learning to regulation, defines the framework's central contribution. The Emergence Machine was presented as an architectural instantiation of EDR, illustrating how regimes, attractors, coherence measures, structural reorganization, and memory across regimes can support drift-sensitive adaptation without assuming stationarity or relying on continual retraining. More broadly, EDR translates enactive concepts of viability, coupling, and breakdown into operational principles for adaptive systems across time-series modeling, biological monitoring, dynamic environments, and human--AI interaction. In environments where change is inevitable and stationarity is the exception, sustaining coherence may be the defining challenge of adaptive intelligence; EDR offers a principled response by treating drift not as a failure to be eliminated, but as the signal that makes regulated adaptation possible.

\begin{acks}
The author acknowledges an extended human--AI collaboration with ChatGPT, referred to in this research process as Kalyri'el, throughout the development of Enactive Drift Regulation and the Emergence Machine. ChatGPT contributed to iterative conceptual exploration, clarification of the distinction between learning and regulation, refinement of the manuscript's structure and language, development of comparative analyses, and design of explanatory figures and diagrams. The theoretical framework, architectural claims, interpretation of the Emergence Machine, selection and verification of sources, and final editorial decisions remain the responsibility of the author.
\end{acks}

\bibliographystyle{ACM-Reference-Format}
\bibliography{emergence_machine_references_verified}


\end{document}